\documentclass[onecolumn,preprintnumbers,amsmath,amssymb,a4paper]{revtex4-2}
\usepackage{bm}
\usepackage{extarrows}
\usepackage{amsfonts}
\usepackage{amsmath}
\usepackage{amssymb}
\usepackage{geometry}
\usepackage{graphicx}
\usepackage{footmisc}
\usepackage{braket}
\usepackage{array}
\usepackage{lipsum}
\usepackage{bibentry,natbib}
\usepackage[colorlinks,citecolor=blue]{hyperref}
\begin{document}

\title{Properties of new even and odd nonlinear coherent states with different parameters}

\author{Cheng Zhang$^{1\dag}
$~\footnotetext{$^\dag$cheng\_ysu@163.com}}
\author{Rui-Jiao Miao$^{2, 1}$}
\author{Xiao-Qiu Qi$^{3\ddagger}$~\footnotetext{$^\ddagger$zjst\_qi@163.com}}

\affiliation{$^1$ Key Laboratory for Microstructural Material Physics of Hebei Province, School of Science, Yanshan University, Qinhuangdao 066004, P. R. China}
\affiliation{$^2$ Laboratory for Biomedical Photonics, Institute of Laser Engineering, Faculty of Materials and Manufacturing, Beijing University of Technology, Beijing 100124, China}
\affiliation{$^3$ Physics Department of Zhejiang Sci-Tech University, Hangzhou 310018, China}
\date{\today}

\begin{abstract}
We construct a class of nonlinear coherent states (NLCSs) by introducing a more general nonlinear function and study their non-classical properties, specifically the second-order correlation function $g^{(2)}(0)$, Mandel parameter $Q$, squeezing, amplitude squared squeezing and Wigner function of the optical field. 
The results indicate that the non-classical properties of the new types of even and odd NLCSs crucially depend on nonlinear functions. More concretely, we find that the new even NLCSs could exhibit the photon-bunching effect whereas the new odd NLCSs could show photon-antibunching effect. The degree of squeezing is also significantly affected by the parameter selection of these NLCSs. By employing various forms of nonlinear functions, it becomes possible to construct NLCSs with diverse properties, thereby providing a theoretical foundation for corresponding experimental investigations.

%By employing various forms of nonlinear functions, it becomes possible to construct NLCSs with diverse properties, thereby providing a theoretical foundation for corresponding experimental investigations.Our construction may find intresting applications in the realm of quantum technology.

\end{abstract}

\maketitle

\section{Introduction}

In 1963, Glauber et al. coined the concept of coherent state~\cite{ref1}, an object ubiquitous in the theoretical formulation of modern quantum optics and a plethora of fascinating research fields, to name a few, ion traps~\cite{ref2,ref3}, cavity QED~\cite{ref4,ref5}, linear optics~\cite{ref6}, superconducting circuits~\cite{ref10}, quantum communication~\cite{ref7,ref8}, quantum precision measurement~\cite{ref9}, etc. The non-classical aspects of coherent states have been the focus of theoretical studies for a long time. 

In recent years, several new classes of coherent states stemming from distinct physical systems have been suggested and utilized to facilitate the investigation of practical issues, for example, nonlinear coherent states~\cite{ref12,ref13,ref14}, photon-added~\cite{ref15,ref16,ref17} and photon-subtracted coherent states~\cite{ref18,ref19,ref20}, Schr$\ddot{o}$dinger cat states~\cite{ref21,ref22}, etc. These states possess a number of non-classical properties that are significantly different from classical coherent states~\cite{ref23,ref24}. In particular, the nonlinear coherent state, defined as the eigenstate of the deformed annihilation operator function $af(N)$ in Ref.~\cite{ref12}, is used to study the non-classical properties of the centroid motion and electromagnetic field that capture the trapped ions~\cite{ref1,ref25,ref26}. In 1997, Sivakumar introduced the concept of even and odd nonlinear coherent states~\cite{ref27}, which are the eigenstates of the operator function $f(N)a^2$. The author discussed the statistical and non-classical properties of the deformed Schr$\ddot{o}$dinger cat states, and found that their non-classical properties rely on the nonlinear function employed in the definition instead of their symmetry. In 2000, Roy et al. introduced a new type of annihilation and generation operator of the so-called $f$ oscillator using the displacement operator. As such, they obtained a new class of nonlinear coherent states and studied their squeezing properties and phase space distribution. Their results showed that this type of nonlinear coherent state exhibits intriguing amplitude-squared squeezing and sub-Poissonian distribution~\cite{ref28}. In 2006, Wang et al. proposed another class of even and odd nonlinear coherent state~\cite{ref28, ref30,ref29} and investigated non-classical properties like squeezing, amplitude-squared squeezing, anti-bunching effect and sub-Poissonian distribution~\cite{ref31,ref32}. Specifically, this class of nonlinear coherent states can show amplitude-squared squeezing in a certain direction in phase space and sub-Poissonian statistics~\cite{ref33,ref34,ref35,ref36}. Furthermore, B. Mojaveri et al. constructed new kinds of nonlinear coherent states by establishing various types of displacement operators which provide rich research in non-classical features such as squeezing, anti-bunching effect and sub-Poissonian statistics as well~\cite{ref37,ref38,ref39,ref40}. 

All of the non-classical properties of the NLCSs discussed in the literature crucially rely on the introduction of some nonlinear functions in the definition of annihilation operators. In this work we extend these considerations further by utilizing more general nonlinear functions and  studying the non-classical properties of the resulting NLCSs. We present a comprehensive investigation of the second-order correlation function $g^{(2)}(0)$, Mandel parameter $Q$, squeezing,  amplitude squared squeezing and Wigner function of these new NLCSs. 

The remaining part of the paper is organized as follows.
The theoretical method is given in Sec.~\ref{sec:II}, where we introduce the nonlinear function we employed and the definitions of the new even and odd NLCSs and their respective second-order correlation function $g^{(2)}(0)$, Mandel parameter $Q$, squeezing, amplitude squared squeezing and Wigner function. The discussion of the non-classical properties of the new NLCSs will be presented in Sec.~\ref{sec:III}.
Finally, a conclusion is given in Sec.~\ref{sec:IV}.

%At present, the appearance of non-classical properties for NLCSs depends only on a relatively simple function form, and thus there is a lack of analysis of non-classical effects exhibited by other nonlinear functions.  This paper will further consider the non-classical effects of the new even and odd NLCSs under a more general nonlinear function. By simulating the second-order correlation function $g^{(2)}(0)$, Mandel parameter $Q$, squeezing and amplitude squared squeezing, the non-classical region of the light field would be more clearly displayed, as well as the non-classical phenomenon would be described more comprehensively.

\section{Theoretical method}\label{sec:II}

New NLCS $\left| \lambda ,f \right\rangle $ is the eigenstate of the operator $(a^+)^{-1}a\frac{1}{f(N)}$ ~\cite{ref29},
\begin{equation}\label{eq:1}
\begin{aligned}
(a^+)^{-1}a\frac{1}{f(N)} \left| \lambda ,f \right\rangle = \lambda \left| \lambda ,f \right\rangle,
\end{aligned}
\end{equation}
where $\lambda$ represents the eigenvalue corresponding to the NLCS, $a^+$ and $a$ are the creation and annihilation operators. The particle number operator $N=a^+a$; $f$ is a nonlinear function related to $N$. Expanding the $\left| \lambda ,f \right\rangle$ in Fock space, one has
\begin{equation}\label{eq:2}
\begin{aligned}
\left| \lambda ,f \right\rangle =\sum\limits_{n=0}^{\infty }{( {{C}_{2n}}\left| 2n \right\rangle +{{C}_{2n+1}}\left| 2n+1 \right\rangle )},
\end{aligned}
\end{equation}
where ${C}_{2n}$ and ${C}_{2n+1}$ are the expansion coefficients. According to the properties of two-photon annihilation operator $(a^+)^{-1}a$,
\begin{equation}\label{eq:3}
\begin{aligned}
(a^+)^{-1}a\left| n \right\rangle ={{\left[ n/(n-1) \right]}^{1/2}}\left| n-2 \right\rangle,
\end{aligned}
\end{equation}
we have
\begin{equation}\label{eq:4}
\begin{aligned}
(a^+)^{-1}a\frac{1}{f(N)} \left| \lambda ,f \right\rangle &= \frac{1}{f(n)}\sum\limits_{n=0}^{\infty}{{{C}_{2n}}\sqrt{\frac{2n}{2n-1}}}\left| 2n-2 \right\rangle+\frac{1}{f(n)}\sum\limits_{n=0}^{\infty }{{{C}_{2n+1}}\sqrt{\frac{2n\text{+}1}{2n}}}\left| 2n-1 \right\rangle.
\end{aligned}
\end{equation}
Using the Eqs.~(\ref{eq:1}), (\ref{eq:2}) and (\ref{eq:4}), ${C}_{2n}$ and ${C}_{2n+1}$ can be given as follows,
\begin{equation}\label{eq:5}
\begin{aligned}
{{C}_{2n}}=f(n)!!\sqrt{\frac{(2n-1)!}{(2n)!!}}{{\lambda }^{n}}{{C}_{0}},
\end{aligned}
\end{equation}
and,
\begin{equation}\label{eq:6}
\begin{aligned}
{{C}_{2n+1}}=f(n)!!\sqrt{\frac{(2n)!!}{(2n+1)!!}}{{\lambda }^{n}}{{C}_{1}}.
\end{aligned}
\end{equation}
When the coefficient ${C}_{2n+1}$ is zero, a new even NLCS is obtained,
\begin{equation}\label{eq:7}
\begin{aligned}
&{{\left| \lambda ,f \right\rangle }_{+}}={{C}_{0}}\sum\limits_{n=0}^{\infty}{\sqrt{\frac{(2n-1)!!}{(2n)!!}}f(n)!!{{\lambda}^{n}}\left| 2n \right\rangle },
\end{aligned}
\end{equation}
where the normalization coefficient ${{\left| {{C}_{0}} \right|}^{{}}}$ is
\begin{equation}\label{eq:8}
\begin{aligned}
{{\left| {{C}_{0}} \right|}^{{}}}=\left[\sum\limits_{n=0}^{\infty}{\frac{\sqrt{2n!}}{{{2}^{n}}n!}f(2n)!!{{\lambda }^{n}}}\right]^{-1}.
\end{aligned}
\end{equation}
Similarly, when the coefficient ${C}_{2n}$ is zero, the new odd NLCS is
\begin{equation}\label{eq:9}
\begin{aligned}
{{\left| \lambda ,f \right\rangle }_{-}}={{C}_{1}}\sum\limits_{n=0}^{\infty }{\sqrt{\frac{(2n)!!}{(2n+1)!!}}f(2n+1)!!{{\lambda }^{n}}\left| 2n+1 \right\rangle }.
\end{aligned}
\end{equation}
The corresponding normalization coefficient $\left| {{C}_{1}} \right|$ is
\begin{equation}\label{eq:10}
\begin{aligned}
{{\left| {{C}_{1}} \right|}^{{}}}=\left[\sum\limits_{n=0}^{\infty}{\frac{{{2}^{n}}n!}{\sqrt{(2n+1)!}}f(2n+1)!!{{\lambda }^{n}}}\right]^{-1}.
\end{aligned}
\end{equation}

This paper mainly discusses the non-classical characteristics of the new NLCSs.
The second-order correlation function $g^{(2)}(0)$ can be used to determine whether a light field is a non-classical light field.
For a single-mode light field, the function $g^{(2)}(0)$ can be expressed as
\begin{equation}\label{eq:11}
\begin{aligned}
{{g}^{(2)}}(0)=\frac{\left\langle {{a}^{+2}}{{a}^{2}} \right\rangle}{{{\left\langle {a^+}a \right\rangle }^{2}}}.
\end{aligned}
\end{equation}
If $g^{(2)}(0)<1$, the light field exhibits a non-classical photon-antibunching effect, which is generally a non-classical light field. For the new even and odd NLCSs given by the equation (\ref{eq:7}) and (\ref{eq:9}), the second-order correlation functions are
\begin{equation}\label{eq:12}
\begin{aligned}
{{g}_{+}}^{(2)}(0)=\frac{\sum\limits_{n=0}^{\infty }{\frac{(2n)!2n(2n-1)}{{{({{2}^{n}}n!)}^{2}}}{{\left[ f(2n)!! \right]}^{2}}}{{\left| \lambda  \right|}^{2n}}}{C_{0}^{2}{{\left[ \sum\limits_{n=0}^{\infty }{\frac{(2n)!2n}{{{({{2}^{n}}n!)}^{2}}}{{\left[ f(2n)!! \right]}^{2}}}{{\left| \lambda  \right|}^{2n}} \right]}^{2}}},
\end{aligned}
\end{equation}
and
\begin{equation}\label{eq:13}
\begin{aligned}
{{g}_{-}}^{( 2 )}( 0 )=\frac{\sum\limits_{n=0}^{\infty }{\frac{{{( {{2}^{n}}n! )}^{2}}}{( 2n-1 )!}{{\left[ f( 2n+1 )!! \right]}^{2}}}{{\left| \lambda  \right|}^{2n}}}{C_{1}^{2}{{\left[ \sum\limits_{n=0}^{\infty }{\frac{{{( {{2}^{n}}n! )}^{2}}}{( 2n )!}{{\left[ f( 2n+1 )!! \right]}^{2}}}{{\left| \lambda  \right|}^{2n}} \right]}^{2}}},
\end{aligned}
\end{equation}
respectively.
The photon statistical characteristic is usually described by the Mandel parameter $Q$ parameter,
\begin{equation}\label{eq:14}
\begin{aligned}
Q=\frac{\left\langle {{N}^{2}} \right\rangle -{{\left\langle N \right\rangle }^{2}}}{\left\langle N \right\rangle }-1.
\end{aligned}
\end{equation}
This parameter mainly describes the deviation for the distribution of photons from the Poisson distribution. And it is another important index to determine whether the light field is a non-classical light field. For $Q<0$, the distribution of photons for this quantum state is narrower than the Poisson distribution, i.e., the sub-Poisson distribution.
For a single-mode light field, the relationship between Mandel parameter $Q$ and $g^{(2)}(0)$ is
\begin{equation}\label{eq:15}
\begin{aligned}
Q=\left\langle N \right\rangle (g^{(2)}(0)-1).
\end{aligned}
\end{equation}
By using the Eqs.(\ref{eq:7}), (\ref{eq:9}) and (\ref{eq:14}), the Mandel parameter $Q$ parameters for the new even and odd NLCSs are
\begin{equation}\label{eq:16}
\begin{aligned}
	{{Q}_{+}}&=\frac{\sum\limits_{n=0}^{\infty}{\frac{(2n)!2n(2n-1)}{{{({{2}^{n}}n!)}^{2}}}{{\left[ f( 2n )!! \right]}^{2}}}{{\left| \lambda  \right|}^{2n}}}{\sum\limits_{n=0}^{\infty}{\frac{(2n)!2n}{{{({{2}^{n}}n!)}^{2}}}{{\left[ f( 2n )!! \right]}^{2}}}{{\left| \lambda  \right|}^{2n}}}-C_{0}^{2}\sum\limits_{n=0}^{\infty}{\frac{(2n)!2n}{{{({{2}^{n}}n!)}^{2}}}{{\left[ f(2n)!! \right]}^{2}}}{{\left| \lambda  \right|}^{2n}}-1,
\end{aligned}
\end{equation}
and
\begin{equation}\label{eq:17}
\begin{aligned}
	{{Q}_{-}}&=\frac{\sum\limits_{n=0}^{\infty }{\frac{{{({{2}^{n}}n!)}^{2}}}{(2n-1)!}{{\left[ f( 2n+1 )!! \right]}^{2}}}{{\left| \lambda  \right|}^{2n}}}{\sum\limits_{n=0}^{\infty }{\frac{{{({{2}^{n}}n!)}^{2}}}{(2n)!}{{\left[ f( 2n+1 )!! \right]}^{2}}}{{\left| \lambda \right|}^{2n}}}-C_{1}^{2}\sum\limits_{n=0}^{\infty }{\frac{{{({{2}^{n}}n!)}^{2}}}{( 2n )!}{{\left[ f(2n+1)!! \right]}^{2}}}{{\left| \lambda  \right|}^{2n}}-1.
\end{aligned}
\end{equation}

%%%%%%%—————start——squeezing——————%%%
The other non-classical properties, such as squeezing and amplitude squared squeezing, of the new odd and even NLCSs can also be investigated. Two orthogonal complex amplitude components, which are two measurable operators of the light field, are defined as~\cite{ref29}
\begin{equation}\label{eq:opera_X}
\begin{split}
	X_1 &= (a+a^{\dagger})/2, \\ 
	X_2 &= (a-a^{\dagger})/{2i}.
\end{split}
\end{equation}
Based on the Hermitian of the two measurable operators, we write the commutation relation and uncertainty relation
\begin{equation}\label{eq:opera_rela}
\begin{gathered}
	\left[X_1,X_2\right] = i/2, \\ 
	(\Delta X_1)^2(\Delta X_2)^2 \geq 1/16.
\end{gathered}
\end{equation}
If a certain orthogonal component of the light field satisfies $(\Delta X_j)^2\leq 1/4$ ($j=1,2$), then there is a squeezing effect on the component of the light field. The degree of the squeezing can be characterized as
\begin{align}
	\label{eq:deg_squeez1}
	D^1(1) &=2\left\langle a^{\dagger}a \right\rangle + \left\langle {a^{\dagger}}^2 + a^2 \right\rangle - {\left\langle a^{\dagger} + a \right\rangle}^2, \\
	\label{eq:deg_squeez2}
	D^2(1) &=2\left\langle a^{\dagger}a \right\rangle - \left\langle {a^{\dagger}}^2 + a^2 \right\rangle + {\left\langle a^{\dagger} + a \right\rangle}^2.
\end{align}
If $D^{j} (1)$ ($j=1,2$) is within the interval $\left[ -1, 0 \right)$, there is a squeezing effect on the $X_j$ component of the light field, and the magnitude of the squeezing effect reflects the degree of squeezing. $D^{j} (1)= -1$ indicates that the light field is completely squeezed on the $X_j$ component.
By substituting Eqs.(\ref{eq:7}) and Eq.(\ref{eq:9}), we can obtain the expectation value of the following operators as
\begin{align}
	\label{eq:exp_a1}
	\langle a \rangle_{\pm} &=\langle a^{\dagger} \rangle_{\pm}=0, \\
	\label{eq:exp_a2}
	\langle a^{\dagger}a \rangle_{+} &=\left| C_0 \right|^{2}\sum\limits_{n=0}^{\infty}\frac{(2n)! 2n}{(2^{n})^2}[f(2n)!!]^2\left| \lambda \right|^{2n}, \\
	\label{eq:exp_a3}
	\langle a^{\dagger}a \rangle_{-} &=\left| C_1 \right|^{2}\sum\limits_{n=0}^{\infty}\frac{(2^{n} n!)^2}{(2n)!}[f(2n+1)!!]^2\left| \lambda \right|^{2n}, \\
	\label{eq:exp_a4}
	\langle a^{2} \rangle_{+} &=\left| C_0 \right|^{2}\sum\limits_{n=0}^{\infty}\frac{(2n+1)! }{(2^{n}n!)^2 }f(2n)!! f(2n+2)!!\left| \lambda \right|^{2n+1}, \\
	\label{eq:exp_a5}
	\langle a^{2} \rangle_{-} &=\left| C_1 \right|^{2}\sum\limits_{n=0}^{\infty}\frac{(2^{n} n!)^{2} (2n+2)}{(2n+1)!}f(2n+1)!! f(2n+3)!! \left| \lambda \right|^{2n+1}. 
\end{align}
%%%%%%%—————end——squeezing——————%%%
%%%%%%%%%%%%%%%%%%%%%%%%%%%%%%%%%
%%%%%----start---amplitude squared squeezing————%%%

Similarly, the two orthogonal complex quadratic components of the light field of the new odd and even NLCSs can be defined as two measurable operators
\begin{equation}\label{eq: opera_Y}
\begin{split}
	Y_1 &= (a^2+{a^{\dagger}}^2)/2, \\ 
	Y_2 &= (a^2-{a^{\dagger}}^2)/{2i},
\end{split}
\end{equation}
which satisfy the following commutation and uncertainty relation
%%%%%
\begin{equation}\label{eq:opera_rela}
\begin{gathered}
	\left[Y_1, Y_2\right] = i(2N+1), \\ 
	(\Delta Y_1)^2(\Delta Y_2)^2 \geq {\left| \left\langle N+ 1/2\right\rangle \right|}^2.
\end{gathered} 
\end{equation}
%%%%%
If a certain orthogonal component $(\Delta Y_j)^2$ $(j=1,2)$ of the light field is less than $\left| \left\langle N+1/2 \right\rangle \right|$, there is an amplitude squared squeezing effect on the $Y_j$ component of the light field. The degree of the amplitude squared squeezing is characterized as
\begin{align}
	\label{eq: deg_squaredsqueez1}
	D^1(2) &=\frac{ 2\left\langle {a^{\dagger}}^{2} a^{2} \right\rangle + \left\langle {a^{\dagger}}^4 + a^4 \right\rangle - {\left\langle {a^{\dagger}}^{2} + a^{2} \right\rangle}^2}{\left\langle a^{2}{a^{\dagger}}^{2} \right\rangle-\left\langle {a^{\dagger}}^{2} a^{2} \right\rangle}, \\
	\label{eq: deg_squaredsqueez2}
	D^2(2) &=\frac{ 2\left\langle {a^{\dagger}}^{2} a^{2} \right\rangle - \left\langle {a^{\dagger}}^4 + a^4 \right\rangle - {\left\langle {a^{\dagger}}^{2} - a^{2} \right\rangle}^2}{\left\langle a^{2}{a^{\dagger}}^{2} \right\rangle-\left\langle {a^{\dagger}}^{2} a^{2} \right\rangle},
\end{align}
where $-1\leq D^j(2) < 0$ represents amplitude squared squeezing effect on the component of $Y_j$ of the light field. The magnitude of $D^j(2)$ reflects the degree of squeezing, where $D^{j} (2)= -1$ indicates that the light field is completely squeezed on the $Y_j$ component. By utilizing Eq.(\ref{eq:7}) and Eq.(\ref{eq:9}), we have
\begin{align}%%%30-35
	\label{eq: exp_squareda1}
	\left\langle {a^{\dagger}}^{2} a^{2} \right\rangle_{+} &=\left| C_0 \right|^{2}\sum\limits_{n=0}^{\infty}\frac{(2n)! 2n (2n-1)}{(2^{n}n!)^2}[f(2n)!!]^2\left| \lambda \right|^{2n}, \\
	\label{eq: exp_squareda2}
	\left\langle {a^{\dagger}}^{2} a^{2} \right\rangle_{-} &=\left| C_1 \right|^{2}\sum\limits_{n=0}^{\infty}\frac{(2^{n} n!)^2}{(2n-1)!}[f(2n+1)!!]^2\left| \lambda \right|^{2n}, \\
	\label{eq: exp_squareda3}
	\left\langle a^{2} {a^{\dagger}}^{2}  \right\rangle_{+} &=\left| C_0 \right|^{2}\sum\limits_{n=0}^{\infty}\frac{(2n+2)! }{(2^{n}n!)^2}[f(2n)!!]^2\left| \lambda \right|^{2n}, \\
	\label{eq: exp_squareda4}
	\left\langle a^{2} {a^{\dagger}}^{2}  \right\rangle_{-}&=\left| C_1 \right|^{2}\sum\limits_{n=0}^{\infty}\frac{(2^{n} n!)^2 (2n+2)(2n+3)}{(2n+1)!}[f(2n+1)!!]^2\left| \lambda \right|^{2n}, \\
	\label{eq: exp_squareda5}
	\langle a^{4} \rangle_{+} &=\left| C_0 \right|^{2}\sum\limits_{n=0}^{\infty}\frac{(2n+3)(2n+1)!}{(2^{n}n!)^2 }f(2n)!! f(2n+4)!!\left| \lambda \right|^{2n+2}, \\
	\label{eq: exp_squareda6}
	\langle a^{4} \rangle_{-} &=\left| C_1 \right|^{2}\sum\limits_{n=0}^{\infty}\frac{(2^{n} n!)^{2} (2n+2)(2n+4)}{(2n+1)!}f(2n+1)!! f(2n+5)!! \left| \lambda \right|^{2n+2}. 
\end{align}
%%%%%----end--amplitude squared squeezing————%%%

Reconstructing and measuring quasi probability distribution function of the quantum states is of great significance as well for studying the evolution process of quantum states so that one could obtain the information and properties of the quantum state during its evolution process. Therefore, we investigate Wigner function which is defined by (\ref{eq:7}) and (\ref{eq:9}). 
The Wigner function is denoted as
%%%
\begin{equation}\label{eq:wigner}
\begin{aligned}
 \Delta(\alpha, \alpha^{\ast})=\int \frac{d^2z}{\pi^2}\left|\alpha+z\right\rangle \left\langle\alpha-z\right|\exp(\alpha z^{\ast}-\alpha^{\ast}z),
\end{aligned}
\end{equation}
%%%
where $\left|\alpha+z\right\rangle$ is a single-mode coherent state with complex $\alpha$. The Wigner operator is then simplified by the technique of integration within an ordered product (IWOP) of operators~\cite{ref41,ref42} and
the Wigner function is denoted as
%%%
\begin{equation}\label{eq:iwop_wigner}
\begin{aligned}
 \Delta(\alpha, \alpha^{\ast})=\frac{2}{\pi}\mathcal{D}(2\alpha)(-)^{N},
\end{aligned}
\end{equation}
%%%
where $N=a^{\dagger}a$, $\mathcal{D}(2\alpha)=\exp(\alpha a^{\dagger}- \alpha^{\ast} a)$ is the translation operator.
The Wigner function of a pure state $\left|\phi\right\rangle$ which can be represented as a superposition state of Fock states $\left| i\right\rangle$ with coefficient $f_i$ is
%%% Wigner of pure state
\begin{equation}\label{eq:pure_wigner}
\begin{aligned}
 W(\alpha,\alpha^{\ast})=\left\langle \phi \right|\Delta(\alpha, \alpha^{\ast})\left| \phi\right\rangle=\frac{2}{\pi}\sum_{i, j=0}^{\infty}f_j^{\ast}f_{i}(-1)^j\chi_{ji}(2\alpha),
\end{aligned}
\end{equation}
where
\begin{equation}\label{eq:chi_wigner}
\chi_{ji}(\alpha)=\left\langle j \right|\mathcal{D}(\alpha)\left| i\right\rangle= \left\{
\begin{aligned}
\sqrt{\frac{i !}{j !}}\alpha^{j-i}L_i^{j-i}(\left|\alpha \right|^2)\exp(-\left|\alpha \right|^2 /2),&  &{j \geq i} \\
\sqrt{\frac{j !}{i !}}{(-\alpha^{\ast})}^{i-j}L_i^{i-j}(\left|\alpha \right|^2)\exp(-\left|\alpha \right|^2 /2),&  &{j < i} 
\end{aligned} \right. 
\end{equation}
$L_n^{\rho}(x)$ is the generalized Laguerre polynomials.

We finally got the Wigner function of the NLCSs by utilizing Eq. (\ref{eq:7}), (\ref{eq:9}), (\ref{eq:pure_wigner}) (\ref{eq:chi_wigner}) as
\begin{equation}\label{eq:nlcs_wigner}
\begin{aligned}
 W_{\pm}(\alpha,\alpha^{\ast})=\frac{2}{\pi}\sum_{m, n=0}^{\infty}(b_{\pm}^{\ast})_{m}(b_{\pm})_{n}(-1)^n\chi_{mn}(2\alpha),
\end{aligned}
\end{equation}
where
\begin{equation}\label{eq:b_wigner}
\begin{aligned}
 (b_{+})_n &=C_{0} \frac{\sqrt{(2n)!}}{2^{n} n!} f(2n)!! \lambda^n, \\
 (b_{-})_n &=C_{1} \frac{2^{n} n!}{\sqrt{(2n+1)!}} f(2n+1)!! \lambda^n.
\end{aligned}
\end{equation}

%%%
%%%%%----end---Wigner Func————%%%

Non-classical properties of the new NLCSs determined by different nonlinear functions $f(n)$ are obviously disparate. Therefore, the following nonlinear functions are taken to study the statistical distribution characteristics of the quantum states.
\begin{equation}\label{eq:18}
\begin{aligned}
	f(n)=\frac{{{L}_{n}}^{i}(\eta^2)}{{ (1+{{n}^{r}}){{L}_{n}}^{j}(\eta^2)}},
\end{aligned}
\end{equation}
where $L_n^m(x)$ is the Laguerre polynomial and $\eta$ is the Lamb-Dicke parameter. By changing $\eta$, as well as the parameters $i$, $j$ and $r$, different forms of nonlinear functions $f(n)$ are obtained, and thus exhibiting different non-classical properties.

\section{Non-classical properties of new even and odd NLCSs}\label{sec:III}

In this section, the different nonlinear functions $f(n)$ are selected to analyze the non-classical characteristics of the new even and odd NLCSs.

%fig1
\begin{figure*}[htbp!]
\includegraphics[scale=0.4]{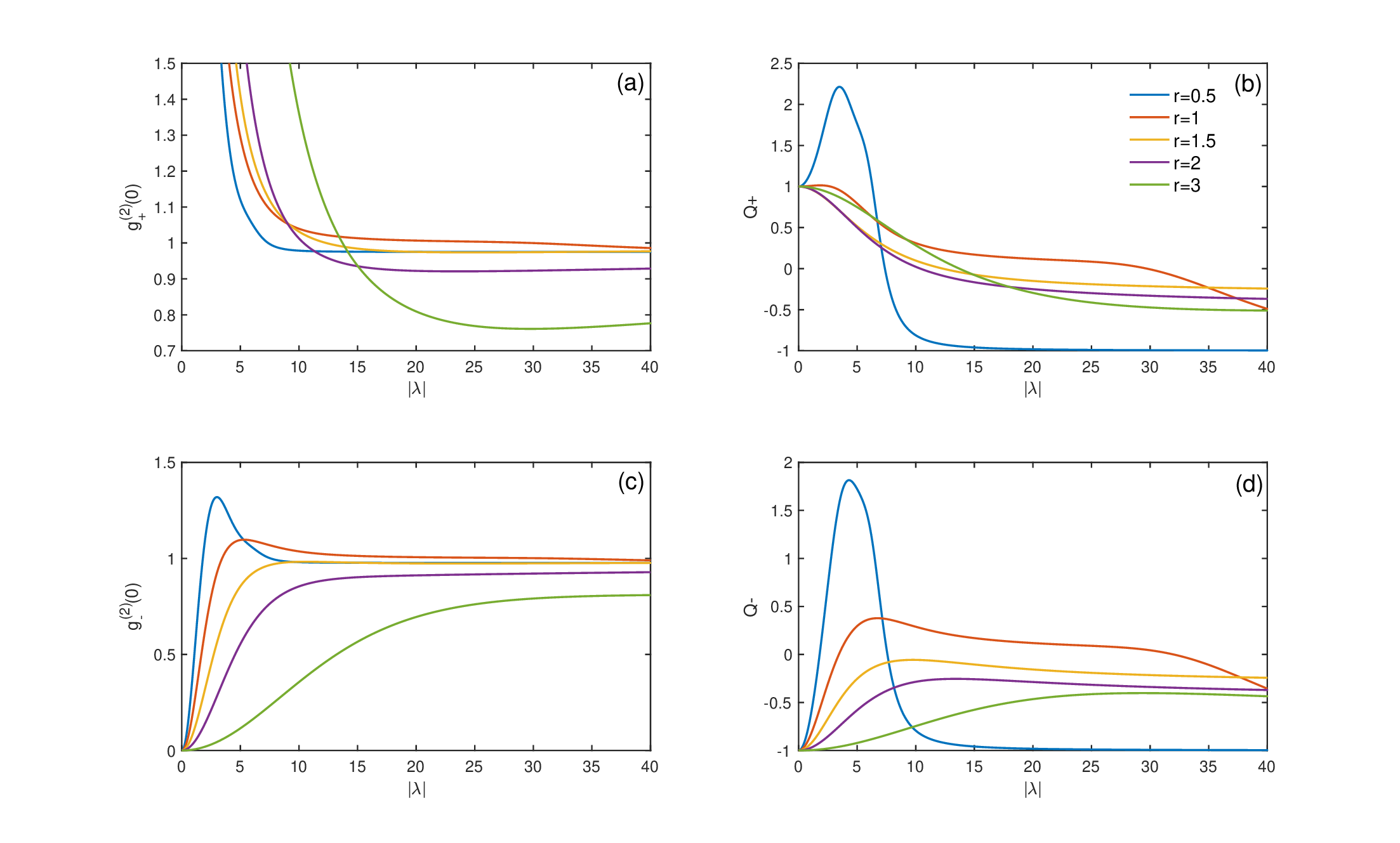}
\caption{The second-order correlation function $g^{(2)}(0)$ and Mandel parameter $Q$ of the new even and odd NLCSs vary with the $|\lambda|$, where (a) and (c) are the curves of even and odd NLCSs for $g^{(2)}(0)$, while (b) and (d) are the curves of Mandel parameter $Q$.}
\label{fig:1}
\end{figure*}

For the case of $i=j$, the nonlinear function in Eq(\ref{eq:18}) degenerates to
\begin{equation}\label{eq:19}
\begin{aligned}
	f(n)=\frac{1}{1+n^{r}}.
\end{aligned}
\end{equation}
Fig.~\ref{fig:1} shows the curves of the second-order correlation function $g^{(2)}(0)$ and Mandel parameter $Q$ varying with $|\lambda|$. The NLCSs will exhibit different non-classical effects due to the changes of $|\lambda|$ and $r$.
With the increase of $|\lambda|$, regardless of the value of $r$, the new even NLCSs change from the photon-bunching effect (super-Poisson distribution) to photon-antibunching effect (sub-Poisson distribution).
Such as the case of $r=2$, the critical position of the sudden change for the characteristic of the NLCS is about $|\lambda|\sim10$.
After this critical value, the NLCS exhibits non-classical characteristics of photon-antibunching effect.
For odd NLCSs, however, the value of $r$ will affect the trend of the second-order correlation function $g^{(2)}(0)$ and the Mandel parameter $Q$.
In particular, when $r=0.5$, the NLCS will change from the photon-antibunching effect to the photon-bunching effect, and then tend to photon-antibunching effect. At $r \geq 1.5$, this phenomenon disappears, i.e., the NLCS will remain in the sub-Poisson distribution.
In addition, as indicated by Eq.(\ref{eq:15}), we observe that when $|\lambda|$ is large, the factor $\left\langle N \right\rangle$ significantly amplifies $g^{(2)}(0)-1$, while for $|\lambda|$ values that are small, it diminishes $g^{(2)}(0)-1$. 
Consequently, in the case of a single-mode light field, the Mandel parameter $Q$ serves as an effective adjustment of the second-order correlation function $g^{(2)}(0)$. This adjustment provides a more intuitive representation of the changes in photon characteristics compared to $g^{(2)}(0)$ alone. This distinction becomes evident when comparing the curves of $g^{(2)}(0)$ and $Q$. Notably, the curves of $g^{(2)}_+(0)$ for the even NLCSs exhibit divergence at $|\lambda|\rightarrow 0$, with all $g^{(2)}(0)$ curves ultimately converging to 1. These phenomena, which do not directly reflect photon characteristics, are absent in the Mandel parameter $Q$. 

%fig2
\begin{figure*}[htbp!]
\includegraphics[scale=0.7]{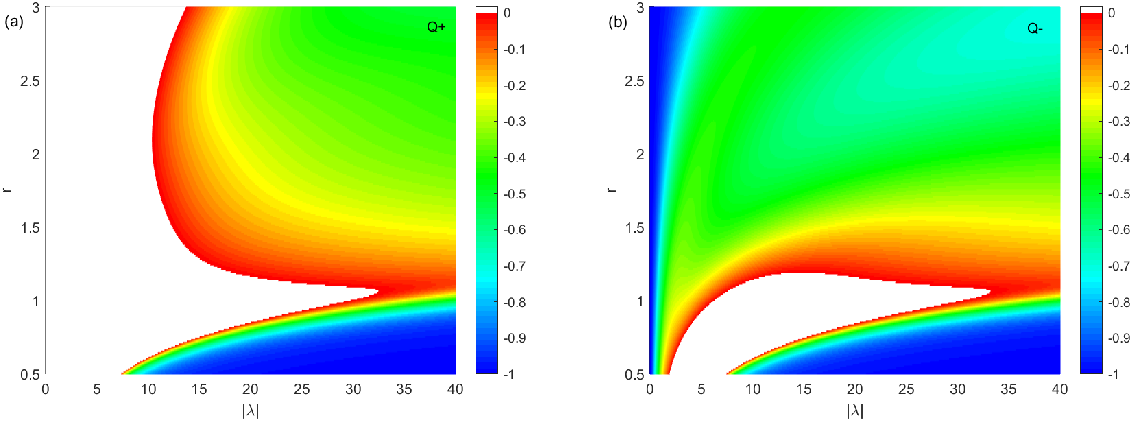}
\caption{The Mandel parameter $Q$ of new even and odd NLCSs vary with the $|\lambda|$ and $r$. (a) and (b) are the results of $Q_+$ and $Q_-$, respectively.}
\label{fig:2}
\end{figure*}

To further analyze the effect of $r$ on the NLCSs, Fig.~\ref{fig:2} shows the Mandel parameter $Q$ changed with the $|\lambda|$  and $r$.
The new odd and even NLCSs show different characteristics when $|\lambda|$ is small. In other words, the even state shows the photon-bunching effect, while the odd state exhibits the photon-antibunching effect.
It can be found that when $|\lambda|$ is in the range of $8\sim35$, the even and odd NLCSs show two transformations: photon-antibunching effect $\rightarrow$ photon-bunching effect $\rightarrow$ photon-antibunching effect. Once the value of $|\lambda|$ is greater than $35$, regardless of the choice of $r$ value, all NLCSs are in the sub-Poisson distribution.

\begin{figure*}[htbp!]
\includegraphics[scale=0.35]{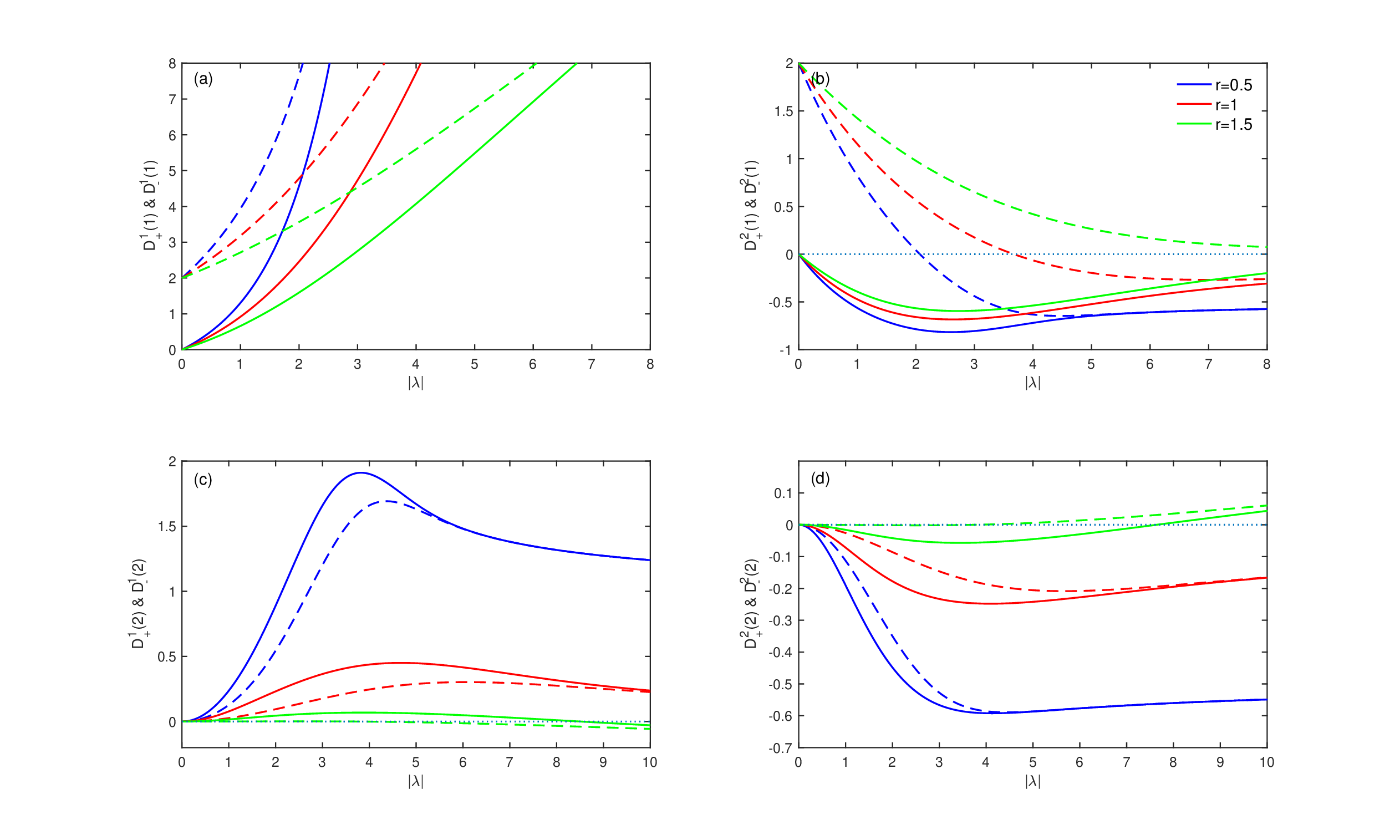}
\caption{The functions $D^l(k)$ ($l,k=1,2$) of the new even (solid line) and odd (dotted line) NLCSs vary with the $|\lambda|$, where the nonlinear function is given in Eq.(\ref{eq:19}).}
\label{figx1}
\end{figure*}

The results of the squeezing degrees $D^l(k)$ ($l,k=1,2$) of the new even and odd NLCSs are shown in Fig.~\ref{figx1}. 
The results indicate that there is a contrasting trend of change between $D^1(k)$ and $D^2(k)$ ($k=1,2$).
For example, the value of D1 is almost positive, meaning there is no significant compression, whereas D2 exhibits compression as $|\lambda|$ changes.
In (a) and (c), only the curve of $D^1(2)$ at $r=1.5$ shows a slight compression.
From (b) and (d), it can be seen that the curve of $D^2_-(1)$ decreases as $|\lambda|$ increases, and eventually stabilizes within the range of $[-1, 0]$, and the degree of compression increases with $r$. 
It is noteworthy that under the same conditions, the $D^l(k)$ function curves of the new even and odd NLCSs will tend to overlap with the increase of $|\lambda|$.

%%%%%%%%%%%%%%%%%%%%%%
%fig3
\begin{figure*}[htbp!]
\includegraphics[scale=0.45]{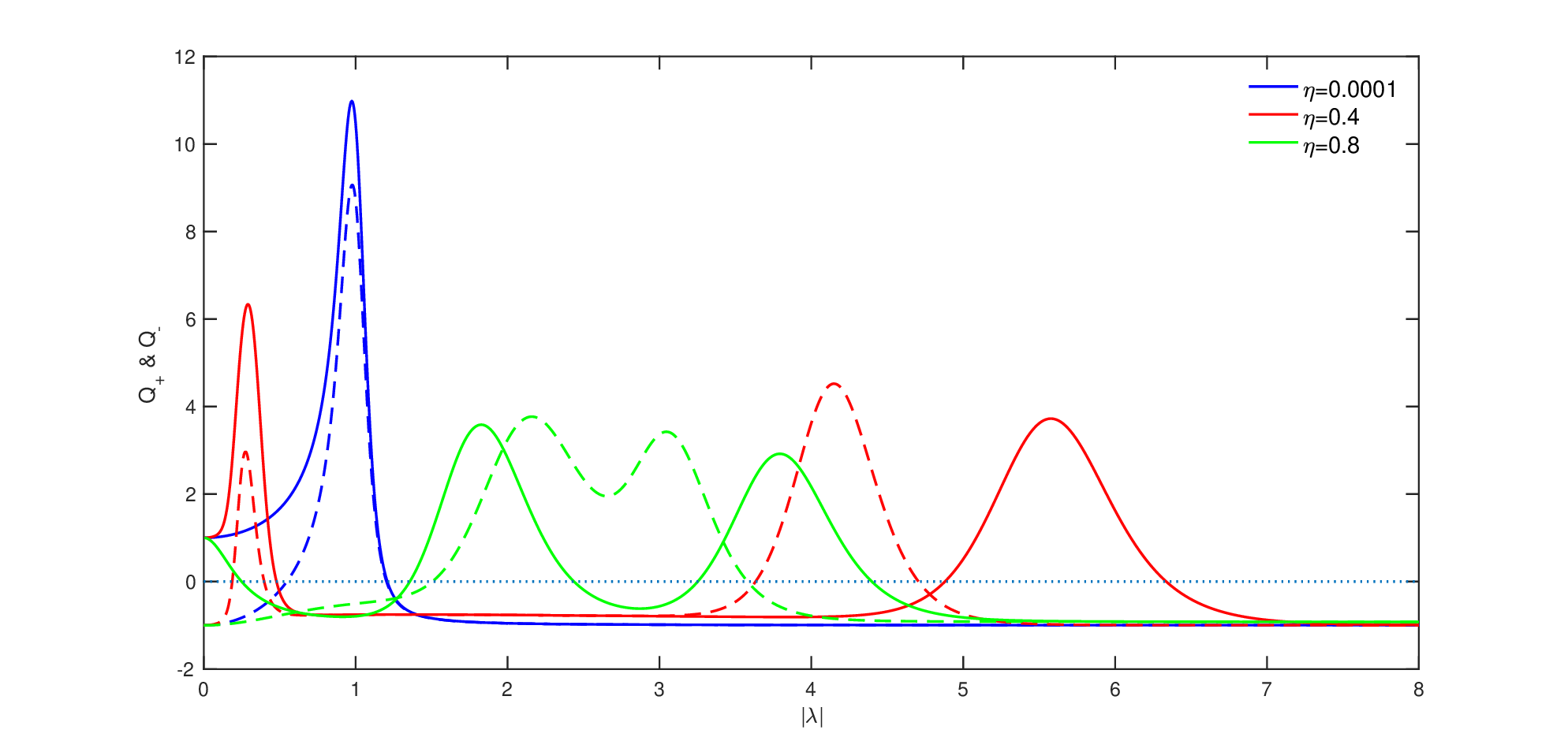}
\caption{The Mandel parameter $Q$ of the new even (solid line) and odd (dotted line) NLCSs vary with the $|\lambda|$.}
\label{fig:3}
\end{figure*}

For the case of $r=1,i=1,j=0$, the nonlinear function is
\begin{equation}\label{eq:20}
\begin{aligned}
	f(n)=\frac{L_{n}^{1}(\eta^{2})}{(1+n) L_{n}^{0}(\eta^{2})}.
\end{aligned}
\end{equation}

Fig.~\ref{fig:3} shows the change curves of the Mandel parameter $Q$ with the $|\lambda|$.
Regardless of the choice of lamb-Dicke parameter $\eta$, these two NLCSs will tend to be sub-Poisson distribution, i.e., the Mandel parameter $Q$ of the new even and odd NLCSs tend to $-1$.
The difference is that the $Q_+=1$ (super-Poisson distribution) for even NLCS at $|\lambda|=0$, while the $Q_-=-1$ (sub-Poisson distribution) for odd NLCS at that position.
The lamb-Dicke parameter $\eta$ has a significant impact on the properties of photons.
It will change the non-classical region of the new even and odd NLCSs, as well as the stability of the Mandel parameter $Q$.
For example, when $\eta$ is large, the oscillations of the curves for the Mandel parameter $Q$ is significantly enhanced, which indicates that the characteristics of the new even and odd NLCSs are unstable with increasing $\eta$, and vice versa.

\begin{figure*}[htbp!]
\includegraphics[scale=0.35]{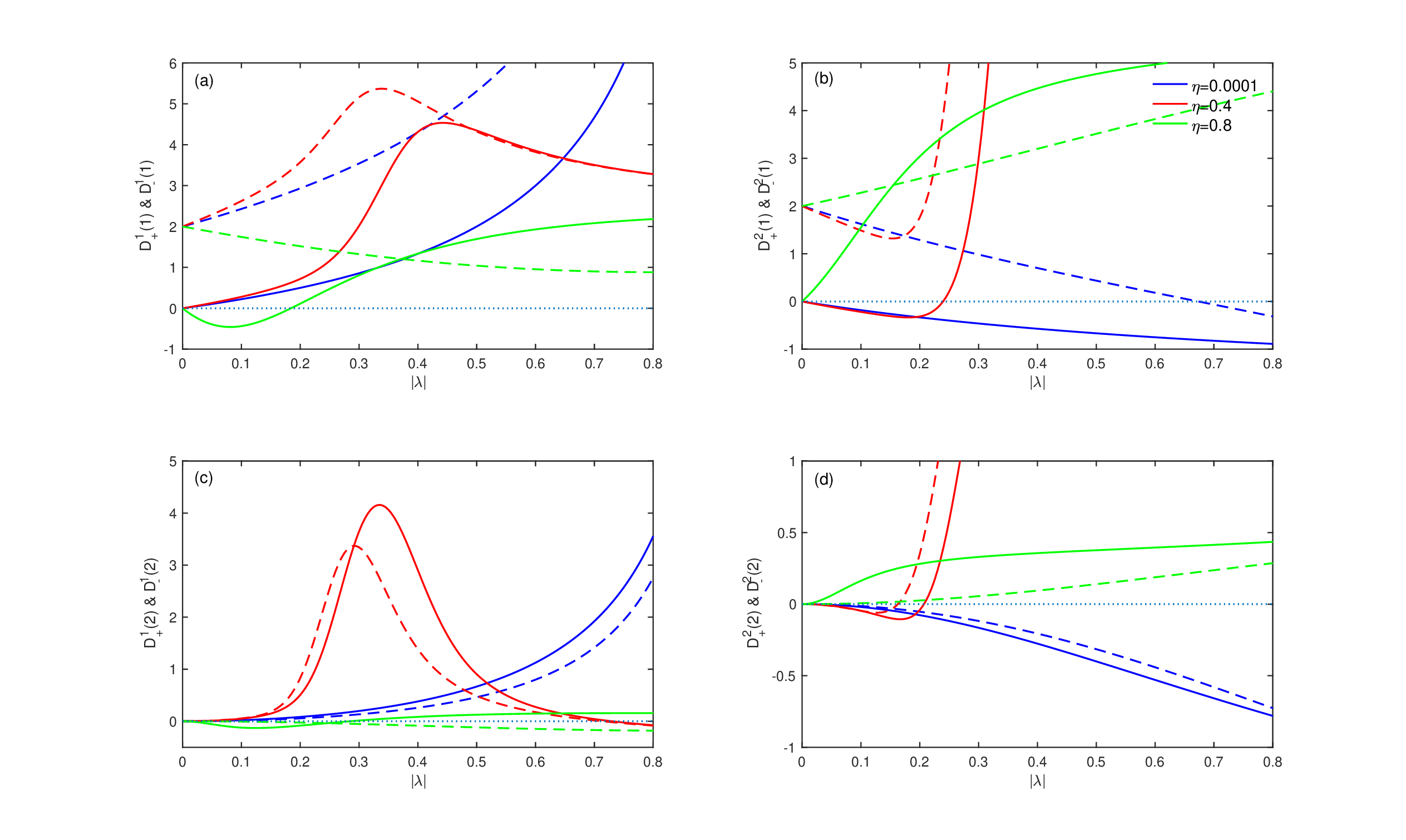}
\caption{The functions $D^l(k)$ ($l,k=1,2$) of the new even (solid line) and odd (dotted line) NLCSs vary with the $|\lambda|$, where the nonlinear function is given in Eq.(\ref{eq:20}).}
\label{figx2}
\end{figure*}

The results of the squeezing degrees $D^l(k)$ ($l,k=1,2$) of the new even and odd NLCSs are shown in Fig.~\ref{figx2}.
The results suggest that different parameters $\eta$ have a significant impact on the variation curves of the $D^l(k)$ functions, with notable fluctuations observed at $\eta=0.4$.
In (a), it can be seen that $D^1_+(1)$ compression only occurs when $\eta=0.8$ and $|\lambda|<0.2$, while this phenomenon does not exhibit in other cases.
For (b), the compression effect is most pronounced at $\eta=0.0001$, where $D^2_-(1)$ showing compression as $|\lambda|$ increases, while $D^2_+(1)$ remains compressed.
From (c) and (d), one can see that the $D^1(2)$ function curves for the new even and odd NLCSs do not show compression at $\eta=0.0001$, whereas the curves of $D^2(2)$ remain compressed.

\begin{figure*}[htbp!]
\includegraphics[scale=0.35]{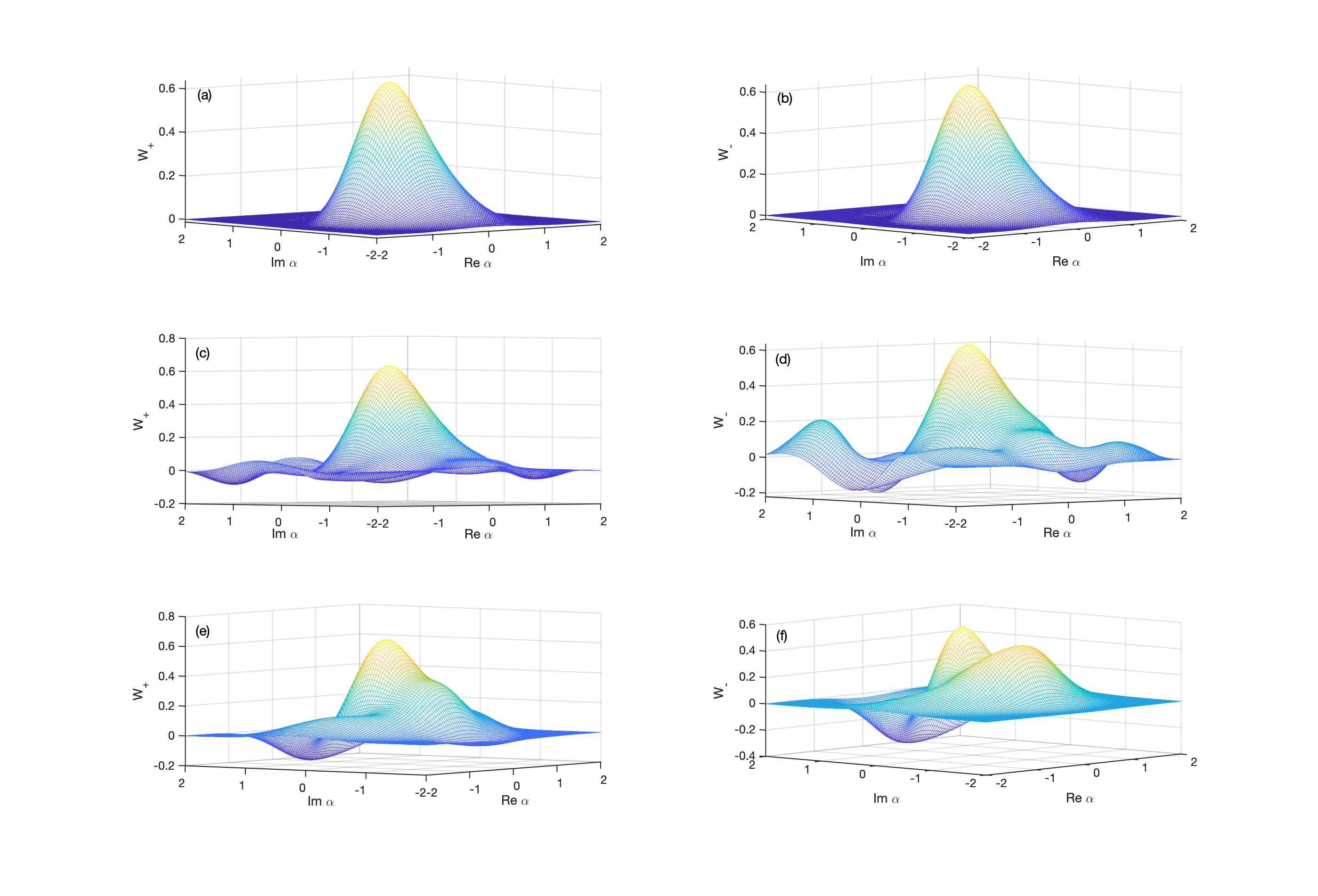}
\caption{The Wigner function of the new even and odd NLCSs vary with the $\alpha=\mathrm{Re} \alpha + i \mathrm{Im} \alpha$, where (a), (c), and (e) are the results of even NLCSs at $\eta=0.2,0.4,0.6$, respectively, while (b), (d), and (f) are the results of odd NLCSs. }
\label{fig5X}
\end{figure*}

Fig.~\ref{fig5X} illustrates how the Wigner functions of new even and odd NLCSs change with respect to the complex parameter $\alpha=\mathrm{Re} \alpha + i \mathrm{Im} \alpha$, with $|\lambda|$ set to $0.1$, and $\eta$ taking values of $0.2$, $0.4$, and $0.6$.
Observing Fig. (a) and (b), it is evident that the Wigner functions of the recently introduced even and odd NLCSs both display a similar unimodal Gaussian-shaped structure when $\eta$ is set to $0.2$.
As the parameter $\eta$ increases, the Wigner function starts to manifest in the negative regions of the phase space, showcasing a downward peak structure. These negative regions signify the emergence of non-classical states, indicative of non-classical phenomena.
Simultaneously, we can observe that the non-classical effects of odd NLCSs are more pronounced compared to even NLCSs when both are evaluated at the same $\eta$.

%%%%%%%%%%%%%%%%%%%%%%
%fig4
\begin{figure*}[htbp!]
\includegraphics[scale=0.4]{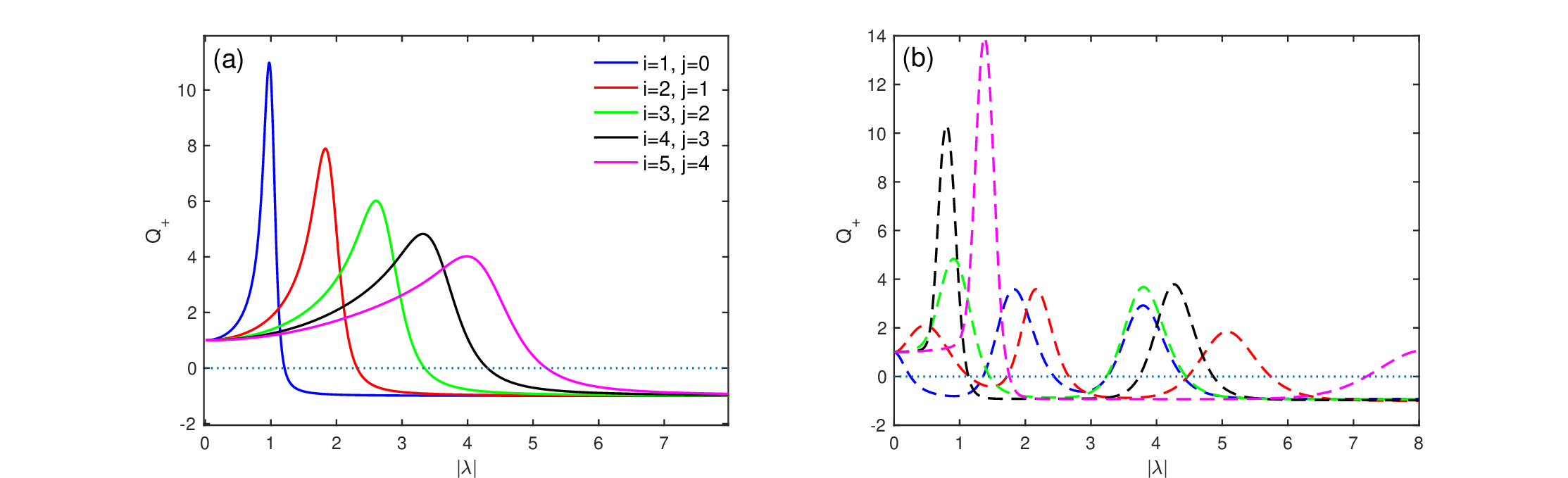}
\caption{The Mandel parameter $Q$ of the new even and odd NLCSs vary with the $|\lambda|$. The curves of $Q$ with $\eta=0.0001$ (solid line) and $0.8$ (dotted line) are shown in (a) and (b), respectively.}
\label{fig:4}
\end{figure*}

In order to analyze the influence of the index $i$, $j$ and the Lamb-Dicke parameter $\eta$ on the photon properties, we consider the case of $r=1$, and thus the nonlinear function is
\begin{equation}\label{eq:21}
\begin{aligned}
	f(n)=\frac{{{L}_{n}}^{i}(\eta^2)}{{ (1+{{n}}){{L}_{n}}^{j}(\eta^2)}},
\end{aligned}
\end{equation}

Fig.~\ref{fig:4} shows the change curves of the Mandel parameter $Q_+$ with $|\lambda|$ for $\eta=0.0001$ and $0.8$. It can be seen from the Fig. (a) that for the case of small $\eta$, the curves of $Q$ generated by different combinations of $i$ and $j$ have the same change trend. The main difference between them is that the critical position from photon-bunching effect to photon-antibunching effect is different, and with the increase of $i$ and $j$, this critical value also increases.
In contrast, for the case of large $\eta$, the above phenomena disappear, and the curves show obvious oscillation between the two effects, as shown in Fig. (b).

\begin{figure*}[htbp!]
\includegraphics[scale=0.5]{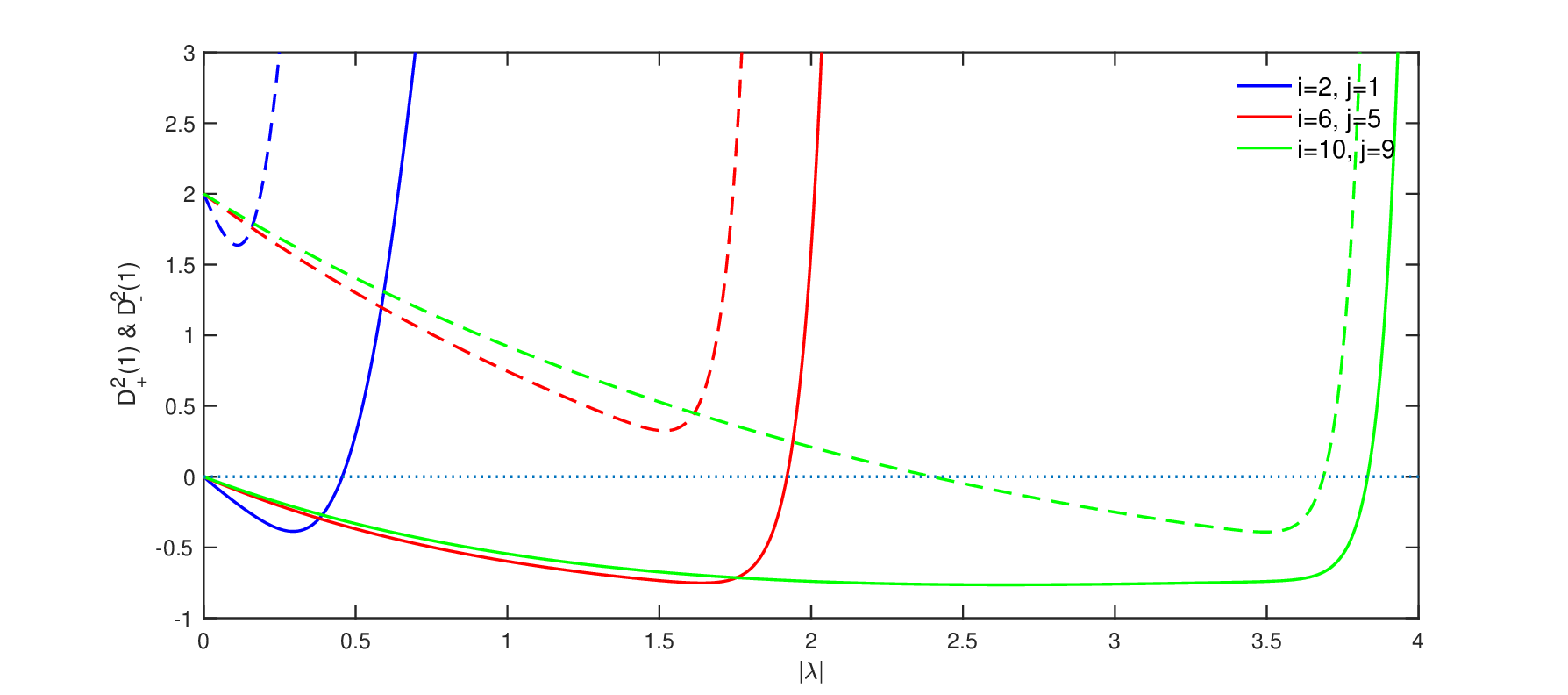}
\caption{The function $D^2(1)$ of the new even (solid line) and odd (dotted line) NLCSs vary with the $|\lambda|$, where the nonlinear function is given in Eq.(\ref{eq:21}).}
\label{figx3}
\end{figure*}

The results of the squeezing degree $D^2(1)$ of the new even and odd NLCSs are shown in Fig.~\ref{figx3}. 
The results show that different $i$ and $j$ do not alter the variation pattern of the $D^2(1)$ function curves, which initially decreases and then increases with changes of $|\lambda|$.
The area where the function curve of $D^2_+(1)$ is less than zero will increase with the increase of $i$ and $j$.
For the $D^2_-(1)$ function, when $i$ and $j$ are large enough, there will be a long decay process as $|\lambda|$ increases, resulting in the appearance of a squeezed region.

\section{Conclusion}\label{sec:IV}

A more general nonlinear function is introduced to construct new even and odd NLCSs. The non-classical properties of the light field for these new even and odd NLCSs, such as the photon-antibunching effect and sub-Poisson distribution, are studied.
Three specific cases, $i=j$, $r=1,i=1,j=0$ and $r=1$, are selected to analyze the impact of different parameters $r$, $\eta$, $i$ and $j$ on the results.
The results show that when some parameters of the nonlinear function are changed, the new even and odd NLCSs exhibit the non-classical effects different from those of the previous works.
When considering the case where $i=j$, it becomes evident that when the parameter $r$ is less than $1$, it significantly influences the changing trends of both the second-order correlation function, denoted as $g^{(2)}(0)$, and the Mandel parameter, denoted as $Q$, with respect to the increase in the magnitude of $|\lambda|$.
In the scenario where $r=1,i=1,j=0$, the Lamb-Dicke parameter $\eta$ plays a crucial role in affecting the stability of the $Q$ curves. Regardless of the chosen value of $\eta$, the Mandel parameter $Q$ eventually exhibits photon antibunching as $|\lambda|$ increases.
For the case of $r=1$, when $\eta$ is small, $Q$ maintains a consistent trend across different combinations of $i$ and $j$.
Furthermore, both the new even and odd NLCSs display intriguing nonclassical properties related to squeezing and amplitude squared squeezing. For instance, when $r=1$, the parameter $X_2$ of the new odd NLCS only exhibits squeezing when $i$ and $j$ are large. Conversely, the $X_2$ parameter of the new even NLCS consistently displays squeezing, with the extent of squeezing increasing as $i$ and $j$ values grow. Additionally, we conducted a comprehensive examination of the Wigner function for these new even and odd NLCSs.
These results can provide theoretical basis for some experiments, such as the preparation of coherent states, etc.

\section*{Acknowledgments}
This research was supported by Startup Fund from Yanshan University, and the Science Foundation of Zhejiang Sci-Tech University under Grant No. 21062349-Y.

 %\end{footnotesize}
%\end{multicols}

\end{document}